\documentclass[12pt,a4paper]{article}
\usepackage[utf8]{inputenc}
\usepackage[T1]{fontenc}
\usepackage[english]{babel}
\usepackage{amsmath,amssymb}
\usepackage{graphicx}
\usepackage{booktabs}
\usepackage[colorlinks=true,linkcolor=blue,citecolor=blue,urlcolor=blue]{hyperref}
\usepackage{xcolor}
\usepackage{multirow}
\usepackage{geometry}

\graphicspath{{figures/}}

\title{Kolmogorov--Arnold Networks as Implicit Regularizers:\\ Noise Robustness and Interpretability for Stellar Classification}
\author{Kristian Sestak\thanks{Independent Researcher. Email: \texttt{kristian.sestak@gmail.com}. ORCID: \href{https://orcid.org/0009-0002-1455-5915}{0009-0002-1455-5915}.}}
\date{}

\begin{document}
\maketitle

\begin{abstract}
\noindent This paper tests whether Kolmogorov--Arnold Networks (KAN~2.0) are genuinely more noise-robust than Multi-Layer Perceptrons (MLP) and XGBoost for stellar classification (star/galaxy/quasar, 100\,000 SDSS~DR17 objects). A naive comparison suggests so: KAN retains $+$9 percentage points over MLP at SNR\,$=$\,5. But equalizing baseline accuracy via weight decay eliminates the gap --- a properly regularized MLP matches KAN to within 1~p.p.\ at all SNR levels, both with and without spectroscopic redshift. The same holds on an independent DESI~DR1 sample with different photometric bands. KAN's robustness thus traces to implicit regularization by $C^2$-smooth B-spline activations, not to architecture. Per-class analysis (20 trials) shows that stars degrade fastest (F1: $0.97 \to 0.75$ at SNR\,$=$\,5), while QSOs remain stable. KAN's native feature importance and SHAP on MLP produce different rankings (Spearman $\rho = -0.37$), capturing complementary aspects of the classification. Colour-index features ($u{-}g$, $g{-}r$, $r{-}i$, $i{-}z$) widen KAN's relative advantage, and a hybrid pipeline routing uncertain MLP predictions to KAN improves low-SNR accuracy. KAN is best understood as a convenient auto-regularizer whose genuine advantage is built-in interpretability.
\end{abstract}


\section{Introduction}
\label{sec:intro}

Classifying astronomical objects from photometric survey data underpins much of modern survey science. The Sloan Digital Sky Survey (SDSS)~\cite{York2000}, the Legacy Survey of Space and Time (LSST)~\cite{Ivezic2019}, and \textit{Euclid}~\cite{Laureijs2011} produce catalogues of billions of sources that require automatic separation into stars, galaxies, and quasars (QSOs) from broadband photometry.

Standard classifiers --- Multi-Layer Perceptrons (MLP)~\cite{Odewahn1992}, Random Forests~\cite{Breiman2001}, gradient-boosted trees~\cite{Chen2016} --- exceed 95\% accuracy on clean benchmarks~\cite{Vasconcellos2011, Kim2017}. The ultraviolet $u$-band, sensitive to stellar temperature and QSO UV excess~\cite{Richards2002}, has anchored colour-based star/galaxy/QSO separation for two decades. Yet these models are trained and evaluated on high-quality data. Faint sources near the detection limit carry photometric uncertainties of 0.1--0.3~mag (SNR\,$\approx$\,3--10), and classifier performance drops accordingly.

Liu et al.~\cite{Liu2024kan} introduced Kolmogorov--Arnold Networks (KAN), replacing fixed nodal activations with learnable B-spline functions on each edge. KAN~2.0~\cite{Liu2024kan2} added sparsification and symbolic regression. Early astrophysical applications exist~\cite{Cui2025, Liu2025gw}, but no controlled test of KAN's noise robustness has been published.

This paper provides such a test on SDSS stellar classification under systematically injected noise. The experimental design goes beyond a naive benchmark: (i)~baseline accuracy is equalized across models to separate regularization from architecture, (ii)~noise-augmented training is tested as a control, (iii)~KAN's interpretability is validated against SHAP, (iv)~colour indices are tested as alternative features, and (v)~a hybrid SNR-adaptive pipeline is proposed. The central finding is that standard weight decay on an MLP replicates KAN's degradation curve; KAN's genuine contribution is that it delivers comparable robustness without tuning, together with inspectable per-feature response functions.

\section{Data}
\label{sec:data}

The primary dataset is the SDSS Data Release~17 stellar classification catalogue~\cite{Abdurrouf2022}: 100\,000 spectroscopically confirmed objects (59\,445 galaxies, 21\,594 stars, 18\,961 QSOs).

Input features are the five SDSS broadband magnitudes ($u$, $g$, $r$, $i$, $z$) and spectroscopic redshift; a subset of experiments drops redshift to test feature dependence. All features are standardized to zero mean and unit variance. The data are split 80/20 with stratified sampling (random seed 42).

For cross-dataset validation (Section~\ref{sec:desi}), a 100\,000-object sample from DESI Data Release~1~\cite{DESI2024} is used, balanced across the three classes, with Legacy Surveys photometry ($g$, $r$, $z$), Gaia BP and RP magnitudes, and spectroscopic redshift. The SDSS sample is class-imbalanced while the DESI sample is balanced; this affects comparability of absolute accuracies between the two.

\section{Methods}
\label{sec:methods}

\subsection{Models}
\label{sec:models}

Six model configurations isolate the effects of architecture, capacity, regularization, and training strategy.

\subsubsection{KAN 2.0}

The Kolmogorov--Arnold representation theorem~\cite{Kolmogorov1957} states that any continuous function $f: [0,1]^n \to \mathbb{R}$ can be decomposed as
\begin{equation}
f(x_1, \dots, x_n) = \sum_{j=0}^{2n} \Phi_j \!\left( \sum_{i=1}^{n} \varphi_{ij}(x_i) \right),
\label{eq:ka}
\end{equation}
where $\varphi_{ij}$ and $\Phi_j$ are univariate functions. KAN~\cite{Liu2024kan} implements this as a network where each edge carries a learnable B-spline of order~$k$:
\begin{equation}
\varphi(x) = \sum_{i} c_i\, B_{i,k}(x),
\label{eq:bspline}
\end{equation}
with learnable coefficients $c_i$ on a grid of $G$ intervals. Cubic B-splines ($k=3$) are $C^2$ continuous.

Our architecture uses width $[6, 24, 12, 3]$ with $G=5$, yielding 8\,148 parameters. Training: Adam, lr\,$=$\,$10^{-3}$, 200 epochs, $L_1$ penalty $\lambda = 0.001$.

\subsubsection{MLP variants}

\textbf{MLP (baseline).} Architecture $[6, 64, 64, 3]$ with SiLU activations~\cite{Elfwing2018}, 4\,803 parameters, Adam lr\,$=$\,$10^{-3}$, 50 epochs, batch size 512. The shorter training schedule relative to KAN (50 vs.\ 200 epochs) was chosen because MLP converges faster; validation loss plateaus by epoch~30.

\textbf{MLP-Large.} Architecture $[6, 128, 128, 3]$, 17\,795 parameters, 200 epochs. Tests whether increased capacity improves robustness.

\textbf{MLP-Reg.} Same as baseline but with weight decay $\lambda_{\mathrm{wd}} = 0.005$, selected to match KAN's clean accuracy of ${\sim}$95.6\%. Tests whether explicit regularization replicates KAN's robustness.

\textbf{MLP-Aug.} Same as baseline, trained for 100 epochs with online noise augmentation: each sample is perturbed with Gaussian noise at random SNR $\in [3, 50]$. Tests whether noise-aware training replicates KAN's robustness.

\subsubsection{XGBoost}

200 gradient-boosted trees, max depth~6, learning rate 0.1, multi-class softmax~\cite{Chen2016}.

\subsection{Noise models}
\label{sec:noise}

Two noise types dominate astronomical photometry and are simulated here.

\textbf{Gaussian (additive) noise} models detector readout noise:
\begin{equation}
\mathbf{x}_{\mathrm{noisy}} = \mathbf{x} + \frac{\|\mathbf{x}\|_2}{\mathrm{SNR}} \cdot \frac{\mathbf{n}}{\|\mathbf{n}\|_2}, \quad \mathbf{n} \sim \mathcal{N}(0, \mathbf{I}).
\end{equation}

\textbf{Poisson (multiplicative) noise} models photon counting:
\begin{equation}
\mathbf{x}_{\mathrm{noisy}} = \mathbf{x} \odot (1 + \boldsymbol{\varepsilon}), \quad \boldsymbol{\varepsilon} \sim \mathcal{N}\!\left(0, \mathrm{SNR}^{-2}\,\mathbf{I}\right).
\end{equation}

SNR levels tested: $\{100, 50, 30, 20, 15, 10, 7, 5, 3, 2, 1\}$. Sections~\ref{sec:naive}--\ref{sec:rdr} use 5 trials per level; Sections~\ref{sec:equal_noredshift}--\ref{sec:desi} use 20 trials for tighter confidence intervals.

\subsection{Metrics}
\label{sec:metrics}

The main metrics are classification accuracy and the \emph{relative degradation rate} (RDR):
\begin{equation}
\mathrm{RDR}(\mathrm{SNR}) = \frac{\mathrm{Acc}_{\mathrm{clean}} - \mathrm{Acc}(\mathrm{SNR})}{\mathrm{Acc}_{\mathrm{clean}}},
\label{eq:rdr}
\end{equation}
which normalizes degradation by each model's own clean accuracy, so that models with different baselines can be compared directly.

\section{Results}
\label{sec:results}

\subsection{Baseline performance}
\label{sec:baseline}

Table~\ref{tab:baseline} lists clean-data performance. MLP reaches the highest accuracy (96.7\%), followed by MLP-Large (96.1\%), KAN and MLP-Aug (${\sim}$95.6\%), MLP-Reg (${\sim}$95.3\%), and XGBoost (94.5\%).

\begin{table}[ht]
\centering
\caption{Baseline performance on clean data (SNR\,$=$\,100).}
\label{tab:baseline}
\begin{tabular}{lcrl}
\toprule
Model & Accuracy & Params & Regularization \\
\midrule
MLP       & 0.967 & 4\,803  & none \\
MLP-Large & 0.961 & 17\,795 & none \\
MLP-Reg   & 0.953 & 4\,803  & weight decay \\
MLP-Aug   & 0.956 & 4\,803  & noise augment. \\
KAN~2.0   & 0.956 & 8\,148  & $L_1$ on splines \\
XGBoost   & 0.945 & 200 trees & tree depth \\
\bottomrule
\end{tabular}
\end{table}

\subsection{Naive comparison: KAN appears superior}
\label{sec:naive}

\begin{figure}[ht]
\centering
\includegraphics[width=\textwidth]{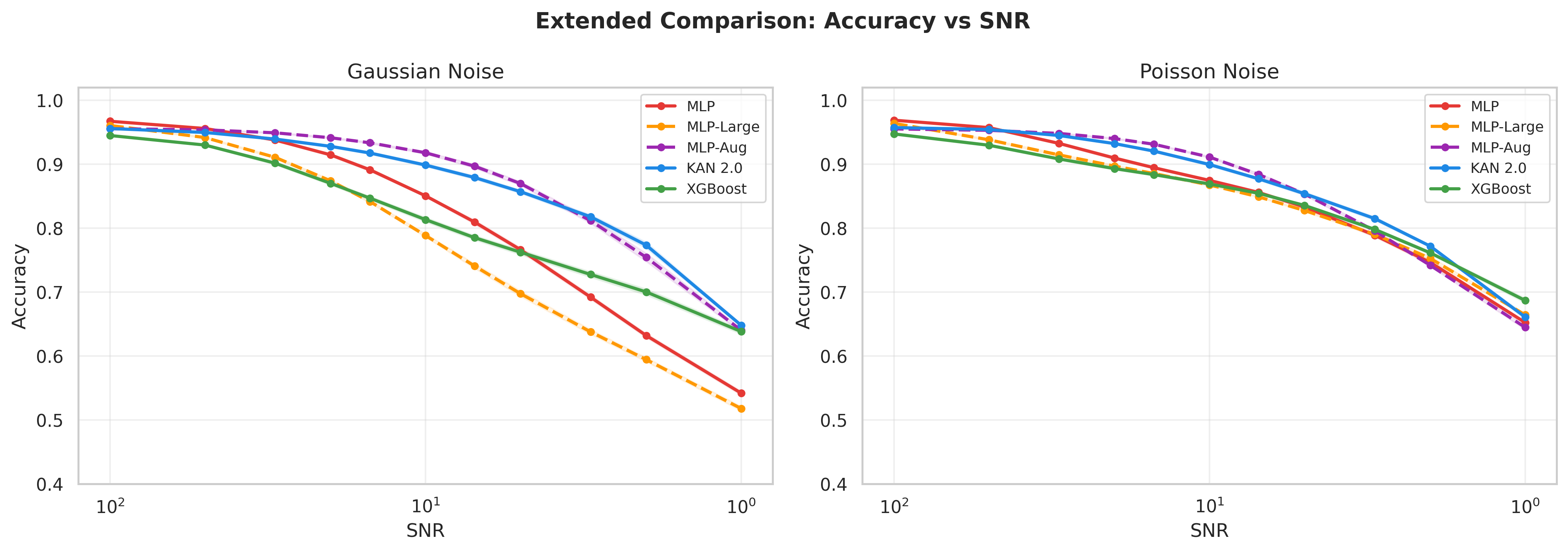}
\caption{Classification accuracy vs.\ SNR for all model configurations under Gaussian (left) and Poisson (right) noise. Shaded regions: $\pm 1\sigma$ over 5 trials. The naive comparison (MLP vs.\ KAN) suggests a large KAN advantage, but MLP-Aug and MLP-Reg (dashed lines) substantially close the gap.}
\label{fig:snr_extended}
\end{figure}

Under default configurations (Fig.~\ref{fig:snr_extended}, Table~\ref{tab:noise}), KAN appears substantially more robust. At SNR\,$=$\,5 it retains 85.7\% accuracy versus 76.6\% for MLP and 76.2\% for XGBoost --- a ${\sim}$9~p.p.\ margin. MLP-Large fares worst (69.7\%): more capacity without smoothness hurts.

\begin{table}[ht]
\centering
\caption{Accuracy (\%) under Gaussian noise. Bold: best at each SNR.}
\label{tab:noise}
\begin{tabular}{rccccc}
\toprule
SNR & KAN & MLP & MLP-Lg & MLP-Aug & XGBoost \\
\midrule
100 & 95.6 & \textbf{96.7} & 96.1 & 95.6 & 94.5 \\
50  & 95.0 & \textbf{95.6} & 94.2 & 95.4 & 93.0 \\
20  & 92.8 & 91.5 & 87.4 & \textbf{94.1} & 87.0 \\
10  & 89.9 & 85.0 & 78.8 & \textbf{91.8} & 81.3 \\
5   & 85.7 & 76.6 & 69.7 & \textbf{87.0} & 76.2 \\
3   & \textbf{81.8} & 69.2 & 63.8 & 81.2 & 72.7 \\
1   & \textbf{64.8} & 54.2 & 51.8 & 64.0 & 63.8 \\
\bottomrule
\end{tabular}
\end{table}

\subsection{Controlled comparison: regularization explains the gap}
\label{sec:controlled}

\begin{figure}[ht]
\centering
\includegraphics[width=0.7\textwidth]{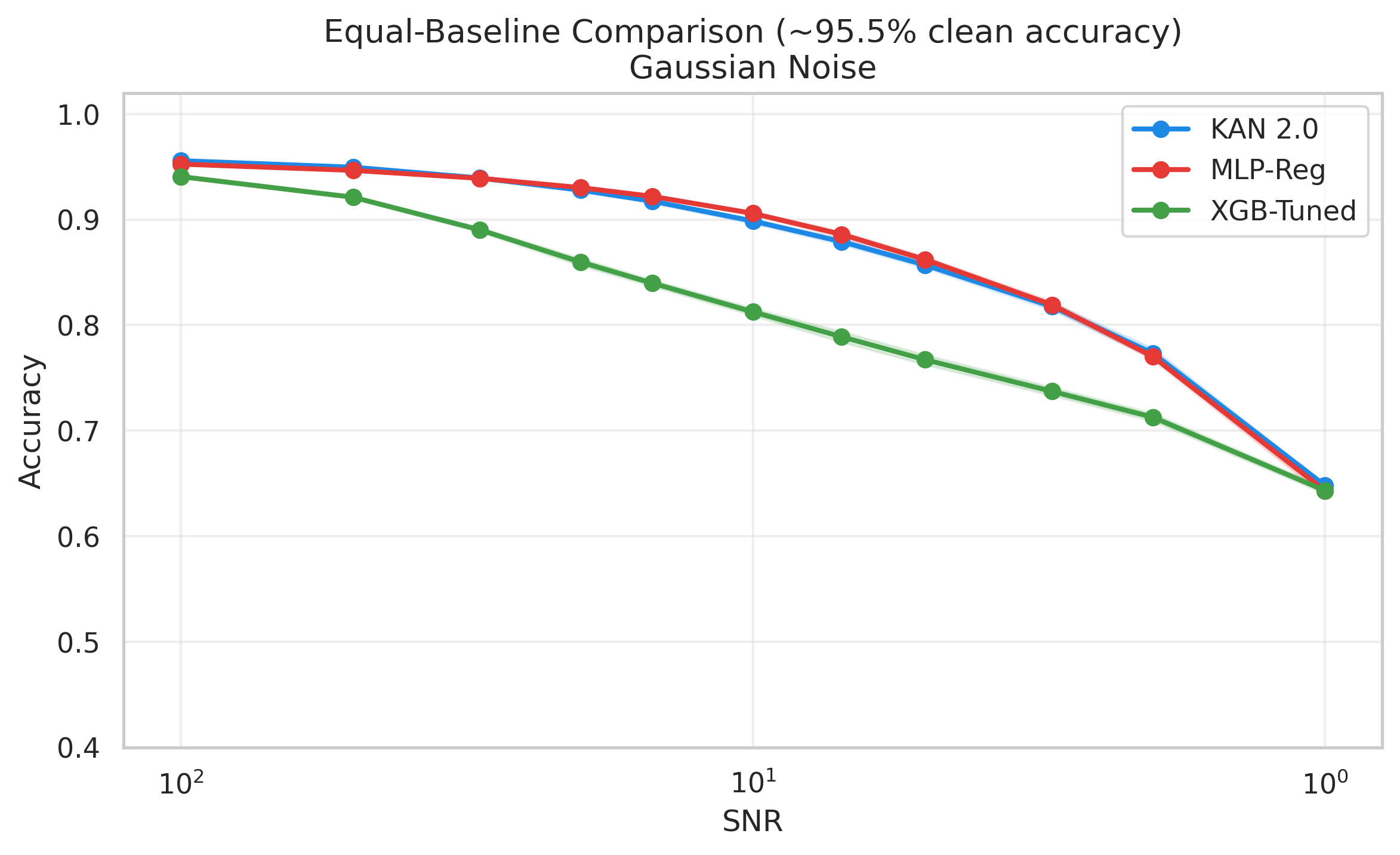}
\caption{Equal-baseline comparison: when MLP is regularized to the same clean accuracy as KAN (${\sim}$95.6\%), the degradation curves are nearly identical, demonstrating that KAN's noise robustness stems from implicit regularization.}
\label{fig:equal_baseline}
\end{figure}

The equal-baseline experiment (Fig.~\ref{fig:equal_baseline}) is the key test. When MLP is regularized via weight decay to match KAN's clean accuracy (${\sim}$95.6\%), the degradation curves become nearly identical (Table~\ref{tab:equal}). At every SNR level the gap between KAN and MLP-Reg stays within 1~p.p. The B-spline basis and $L_1$ penalty constrain KAN to roughly the same degree as explicit weight decay constrains MLP.

XGBoost could not be tuned to the 95.6\% target (best: 94.1\% with 400 trees, depth~10); it is included for completeness.

\begin{table}[ht]
\centering
\caption{Equal-baseline comparison (Gaussian noise). KAN and MLP-Reg start at ${\sim}$95.6\% clean accuracy. XGBoost-Tuned could not reach this target (best: 94.1\%).}
\label{tab:equal}
\begin{tabular}{rccc}
\toprule
SNR & KAN~2.0 & MLP-Reg & XGB-Tuned$^\dagger$ \\
\midrule
100 & 95.6 & 95.3 & 94.1 \\
50  & 95.0 & 94.7 & 92.1 \\
20  & 92.8 & 93.0 & 86.0 \\
10  & 89.9 & \textbf{90.6} & 81.3 \\
5   & 85.7 & \textbf{86.2} & 76.7 \\
3   & 81.8 & \textbf{81.9} & 73.7 \\
1   & \textbf{64.8} & 64.3 & 64.3 \\
\bottomrule
\multicolumn{4}{l}{\small $^\dagger$ Could not be tuned to 95.6\% target.}
\end{tabular}
\end{table}

\subsection{Relative degradation rate}
\label{sec:rdr}

\begin{figure}[ht]
\centering
\includegraphics[width=\textwidth]{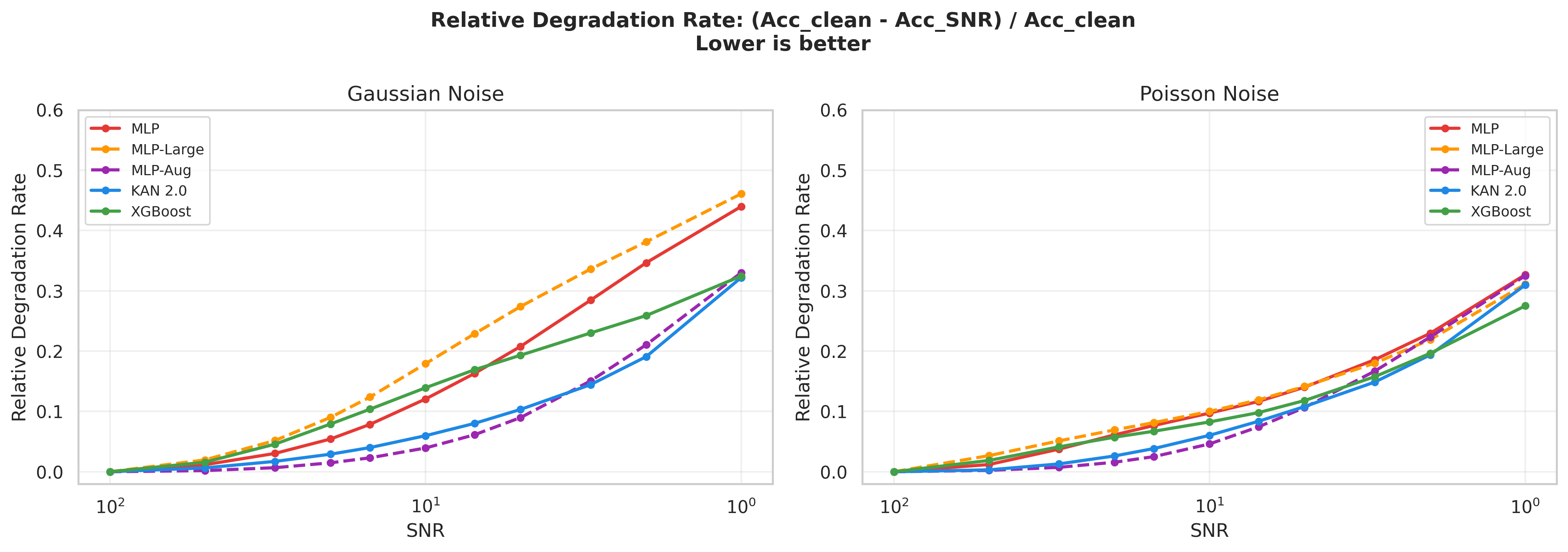}
\caption{Relative degradation rate (lower is better). KAN~2.0 and MLP-Aug show similar degradation profiles, confirming that KAN's robustness is equivalent to what noise augmentation provides.}
\label{fig:rdr}
\end{figure}

Fig.~\ref{fig:rdr} plots the RDR (equation~\ref{eq:rdr}). The default MLP degrades fastest (RDR\,$=$\,20.8\% at SNR\,$=$\,5), and MLP-Large worst of all (27.4\%). KAN (10.3\%) and MLP-Aug (9.0\%) track each other closely.

\subsection{Interpretability: KAN's unique advantage}
\label{sec:interpretability}

KAN's noise robustness can be replicated by regularization; its \emph{interpretability} cannot. The edge-based architecture provides two forms of analysis absent from MLPs.

\begin{figure}[ht]
\centering
\includegraphics[width=0.7\textwidth]{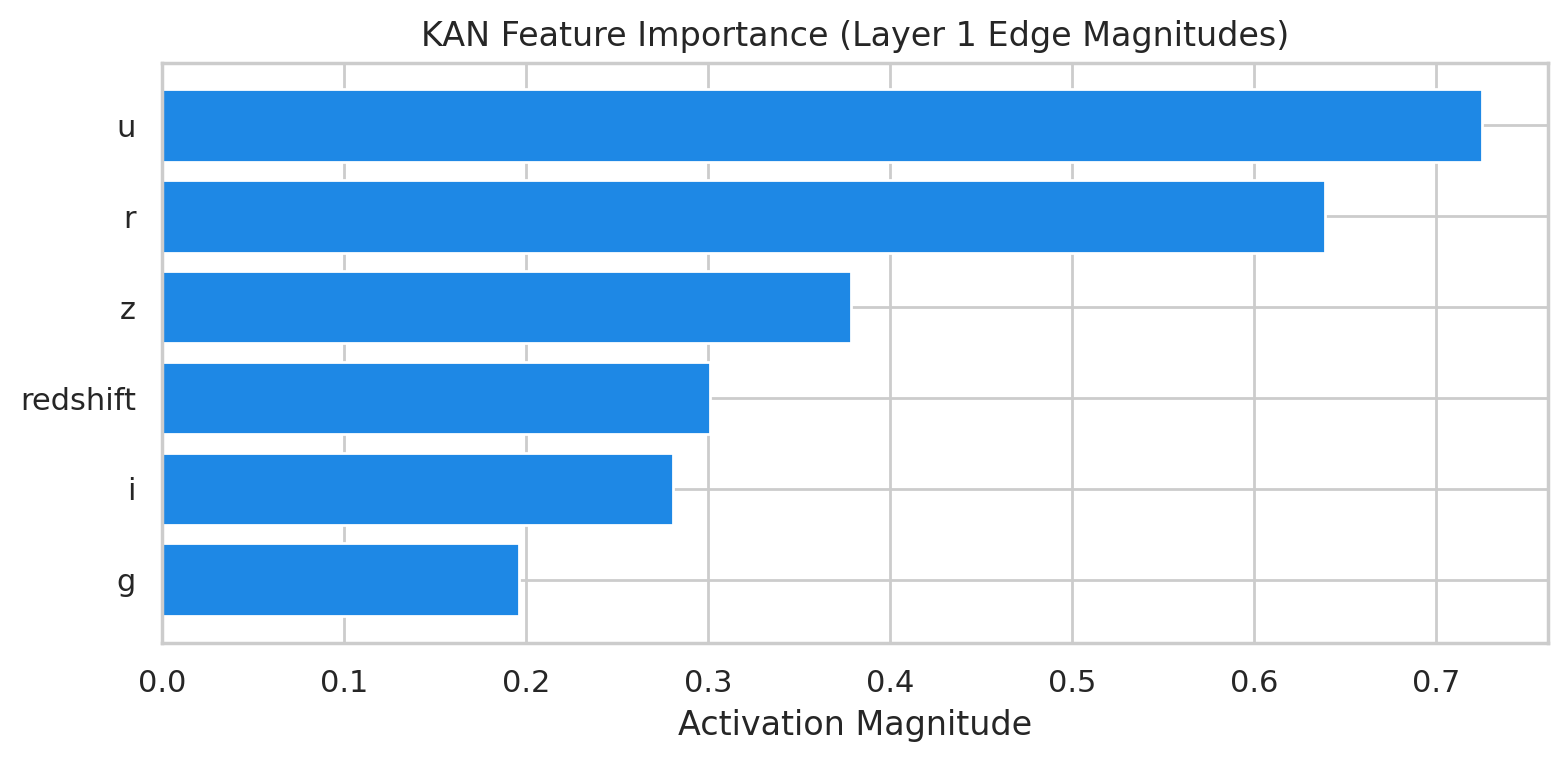}
\caption{KAN feature importance from first-layer edge activation magnitudes. The $u$-band dominates, consistent with its known sensitivity to stellar temperature and QSO UV excess~\cite{Richards2002}.}
\label{fig:feature_importance}
\end{figure}

\textbf{Feature importance.} First-layer edge activation magnitudes rank input features by their contribution to classification (Fig.~\ref{fig:feature_importance}). KAN places the $u$-band first (activation magnitude 0.72), then $r$ (0.63), $z$ (0.38), redshift (0.30), $i$ (0.28), and $g$ (0.19). The $u$-band dominance is expected: it is sensitive to stellar temperature, metallicity, and QSO UV excess~\cite{Richards2002}. The ranking comes directly from the learned network structure, with no post-hoc computation.

\begin{figure}[ht]
\centering
\includegraphics[width=\textwidth]{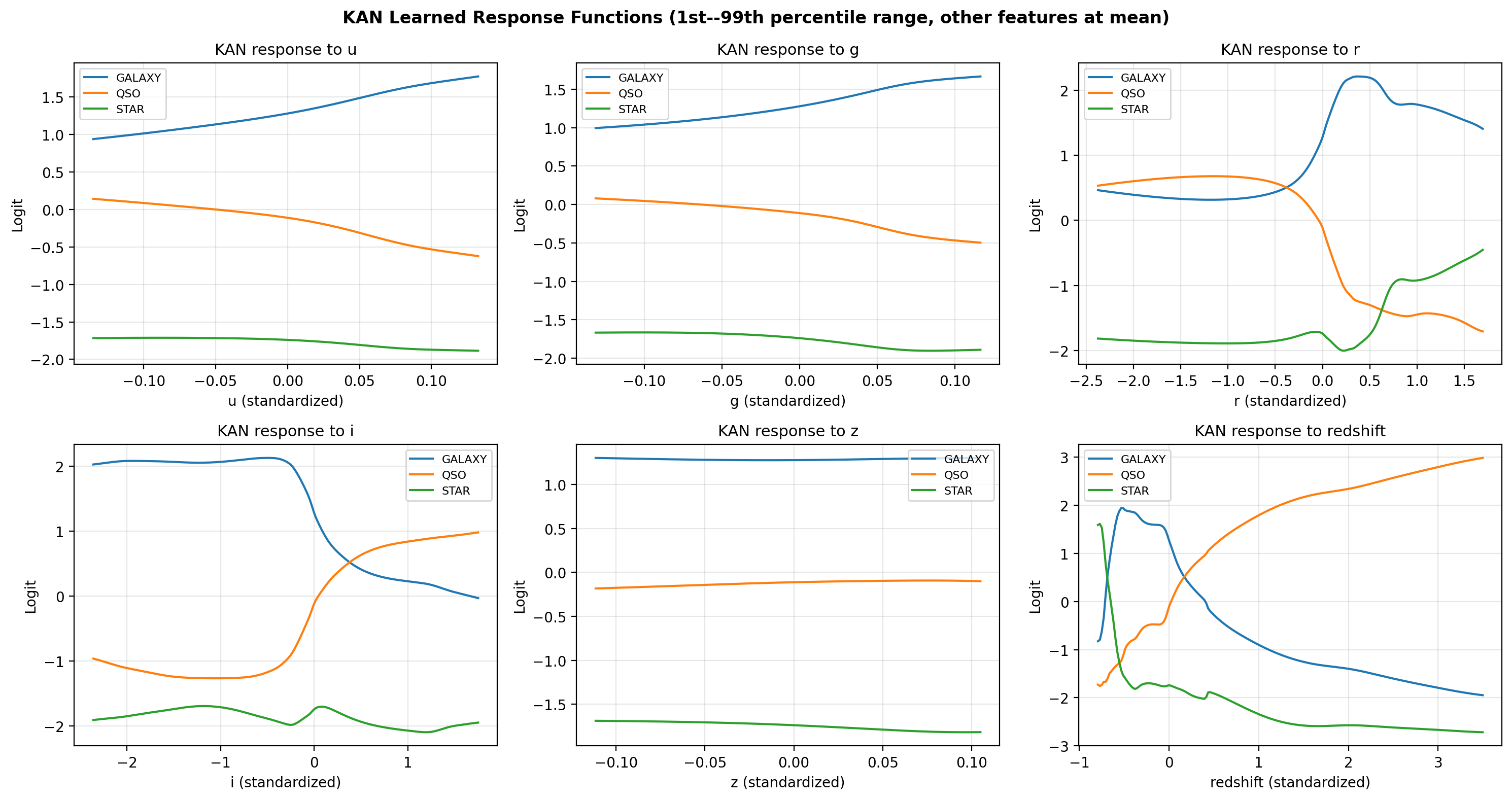}
\caption{KAN learned response functions: output logit per class as a function of a single input feature (other features at mean, 1st--99th percentile range). The redshift response cleanly separates stars from extragalactic objects; the $u$-band response distinguishes the QSO population.}
\label{fig:response}
\end{figure}

\textbf{Learned response functions.} Each KAN edge learns a univariate spline mapping an input feature to a class logit. Fig.~\ref{fig:response} shows these functions over the 1st--99th percentile range (other features at their mean). The redshift response exhibits a sharp transition near $z \approx 0$, cleanly separating stars from extragalactic objects. The $u$-band response shows non-linear regimes corresponding to the blue QSO population. These curves are directly inspectable; extracting comparable information from an MLP requires approximate, computationally expensive post-hoc methods (SHAP, LIME).

\subsection{Photometry-only classification}
\label{sec:noredshift}

\begin{figure}[ht]
\centering
\includegraphics[width=\textwidth]{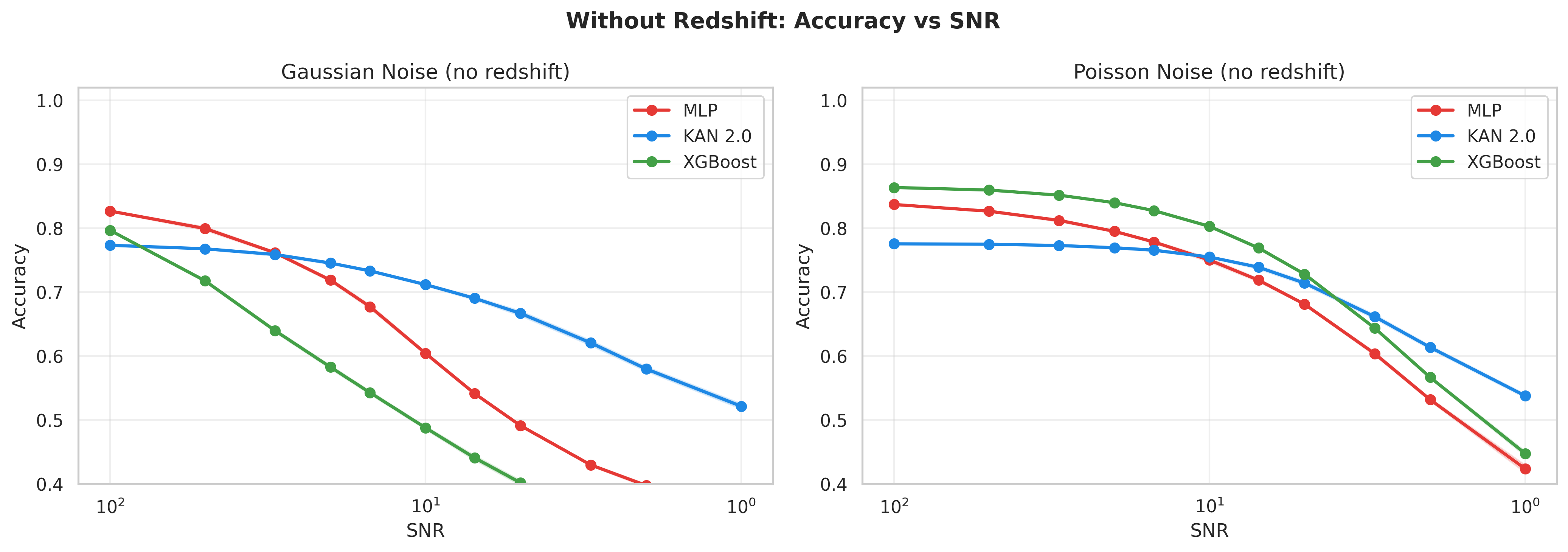}
\caption{Photometry-only classification (no spectroscopic redshift). KAN degrades more slowly but starts from a lower baseline, reflecting stronger implicit regularization.}
\label{fig:snr_noredshift}
\end{figure}

Dropping spectroscopic redshift (a strong prior) and retaining only the five broadband magnitudes reduces baseline accuracies substantially (MLP: 82.7\%, XGBoost: 79.7\%, KAN: 77.3\%). KAN's lower baseline points to stronger implicit regularization on this harder task. Under noise, KAN's relative advantage widens: at SNR\,$=$\,5 it retains 66.7\% versus 49.1\% for MLP (Fig.~\ref{fig:snr_noredshift}).

KAN's 77.3\% baseline suggests underfitting; part of its slower degradation follows from a smoother, less detailed decision boundary.

\subsection{Equal-baseline without redshift}
\label{sec:equal_noredshift}

\begin{figure}[ht]
\centering
\includegraphics[width=0.7\textwidth]{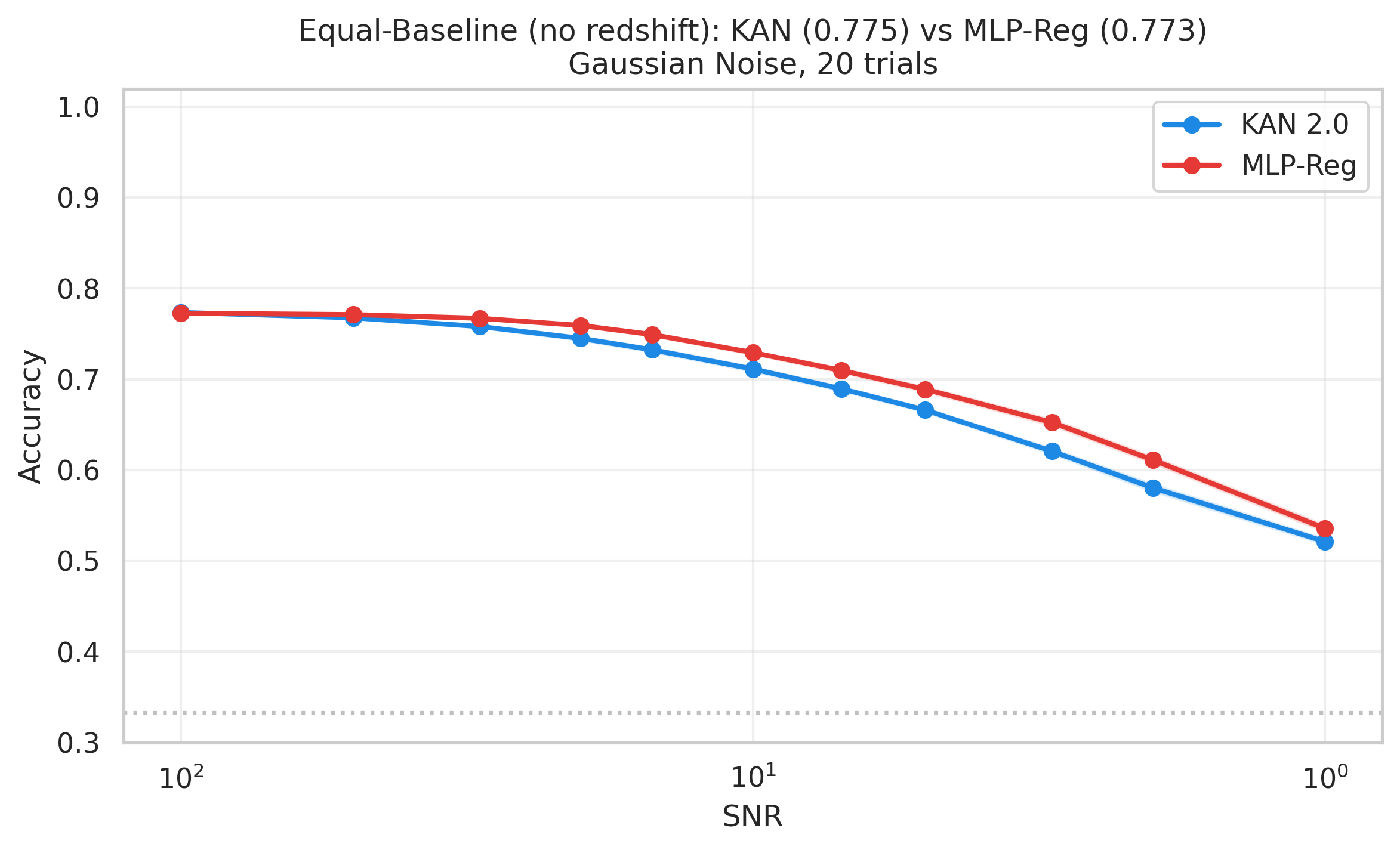}
\caption{Equal-baseline comparison without redshift (20 trials). MLP-Reg slightly outperforms KAN, confirming that the regularization explanation holds in the photometry-only setting.}
\label{fig:equal_noz}
\end{figure}

To address the underfitting concern, the equal-baseline experiment is repeated in the photometry-only setting. MLP-Reg (weight decay\,$=$\,0.008) is tuned to match KAN's 77.3\% baseline. Table~\ref{tab:equal_noz} and Fig.~\ref{fig:equal_noz} show MLP-Reg matching or slightly exceeding KAN at all SNR levels --- the regularization explanation holds even when KAN underfits.

\begin{table}[ht]
\centering
\caption{Equal-baseline comparison without redshift (Gaussian noise, 20 trials). MLP-Reg tuned to match KAN's 77.3\% baseline.}
\label{tab:equal_noz}
\begin{tabular}{rcc}
\toprule
SNR & KAN~2.0 & MLP-Reg \\
\midrule
100 & 77.4 & 77.3 \\
50  & 76.8 & 77.1 \\
20  & 74.5 & 75.9 \\
10  & 71.1 & 72.9 \\
5   & 66.6 & \textbf{68.9} \\
3   & 62.1 & \textbf{65.2} \\
1   & 52.1 & \textbf{53.6} \\
\bottomrule
\end{tabular}
\end{table}

\subsection{Per-class degradation}
\label{sec:perclass}

\begin{figure}[ht]
\centering
\includegraphics[width=\textwidth]{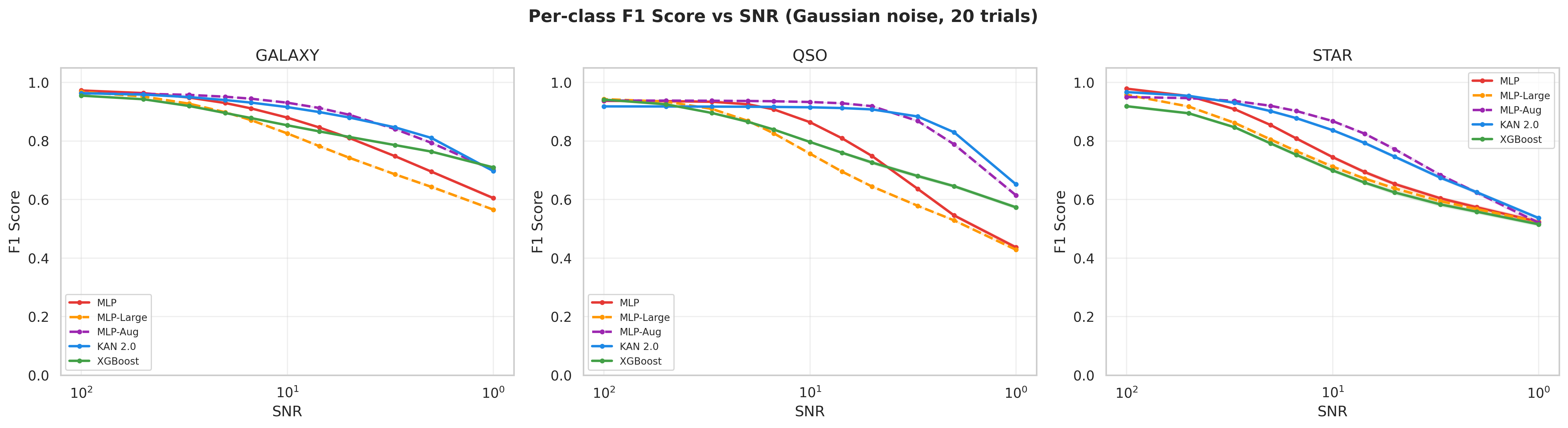}
\caption{Per-class F1 score vs.\ SNR under Gaussian noise (20 trials). Stars degrade fastest; QSOs are most robust. Shaded regions: $\pm 1\sigma$.}
\label{fig:perclass_f1}
\end{figure}

Per-class F1 scores (Fig.~\ref{fig:perclass_f1}, Table~\ref{tab:perclass}) reveal unequal degradation across object types. Stars degrade fastest (F1: $0.97 \to 0.75$ at SNR\,$=$\,5 for KAN) --- stellar classification depends on precise colour ratios that are particularly noise-sensitive. QSOs are the most stable (F1\,$=$\,0.91 at SNR\,$=$\,5), as their distinctive spectral energy distributions provide redundant cues. Galaxies fall between the two (F1\,$=$\,0.88).

\begin{table}[ht]
\centering
\caption{Per-class F1 scores at key SNR levels (Gaussian noise, 20 trials). KAN and MLP-Aug results shown.}
\label{tab:perclass}
\begin{tabular}{rlccc}
\toprule
SNR & Model & Galaxy & QSO & Star \\
\midrule
\multirow{2}{*}{100} & KAN    & 0.963 & 0.918 & 0.967 \\
                     & MLP-Aug & 0.963 & 0.938 & 0.950 \\
\multirow{2}{*}{5}   & KAN    & 0.880 & 0.908 & 0.746 \\
                     & MLP-Aug & 0.889 & 0.919 & 0.772 \\
\multirow{2}{*}{3}   & KAN    & 0.847 & 0.884 & 0.675 \\
                     & MLP-Aug & 0.841 & 0.868 & 0.684 \\
\bottomrule
\end{tabular}
\end{table}

\subsection{Colour-index features}
\label{sec:colors}

\begin{figure}[ht]
\centering
\includegraphics[width=0.7\textwidth]{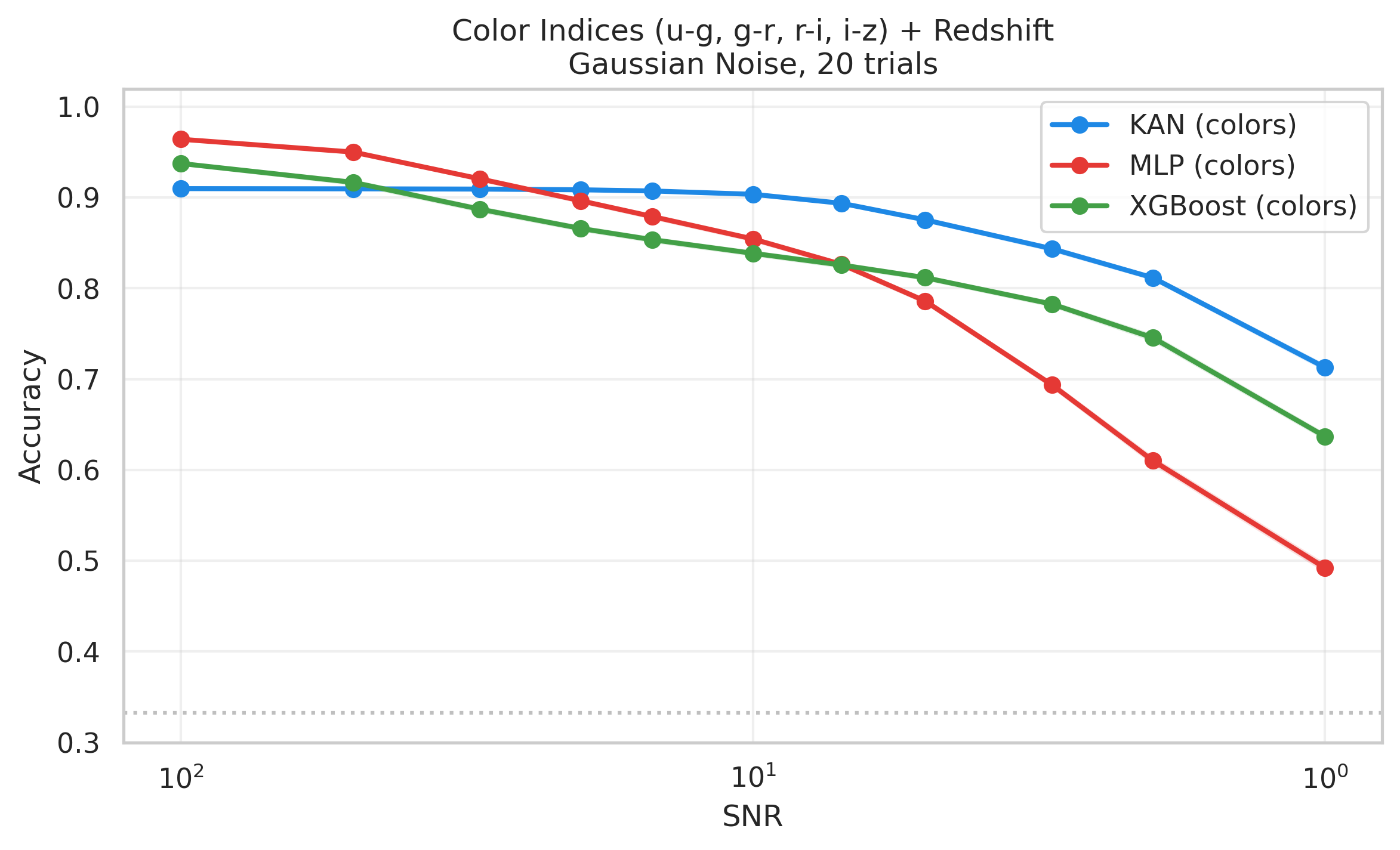}
\caption{Noise robustness with colour-index features ($u{-}g$, $g{-}r$, $r{-}i$, $i{-}z$, redshift). KAN's smooth activations provide even stronger relative robustness on colour indices than on raw magnitudes.}
\label{fig:color_indices}
\end{figure}

Replacing raw magnitudes with colour indices ($u{-}g$, $g{-}r$, $r{-}i$, $i{-}z$) plus redshift changes the picture (Fig.~\ref{fig:color_indices}). KAN's clean accuracy drops to 91.0\% (MLP: 96.4\%), but it degrades far more slowly: at SNR\,$=$\,5 KAN retains 87.6\% versus 78.6\% for MLP, a ${\sim}$13~p.p.\ gap compared to ${\sim}$9~p.p.\ on raw magnitudes. B-spline activations appear particularly effective on features that are already physically meaningful ratios. No equal-baseline experiment was performed for this setting, however; the 5.4~p.p.\ baseline gap is larger than in the raw-magnitude case, and part of the advantage may again stem from implicit regularization.

\subsection{SHAP validation of KAN feature importance}
\label{sec:shap}

\begin{figure}[ht]
\centering
\includegraphics[width=\textwidth]{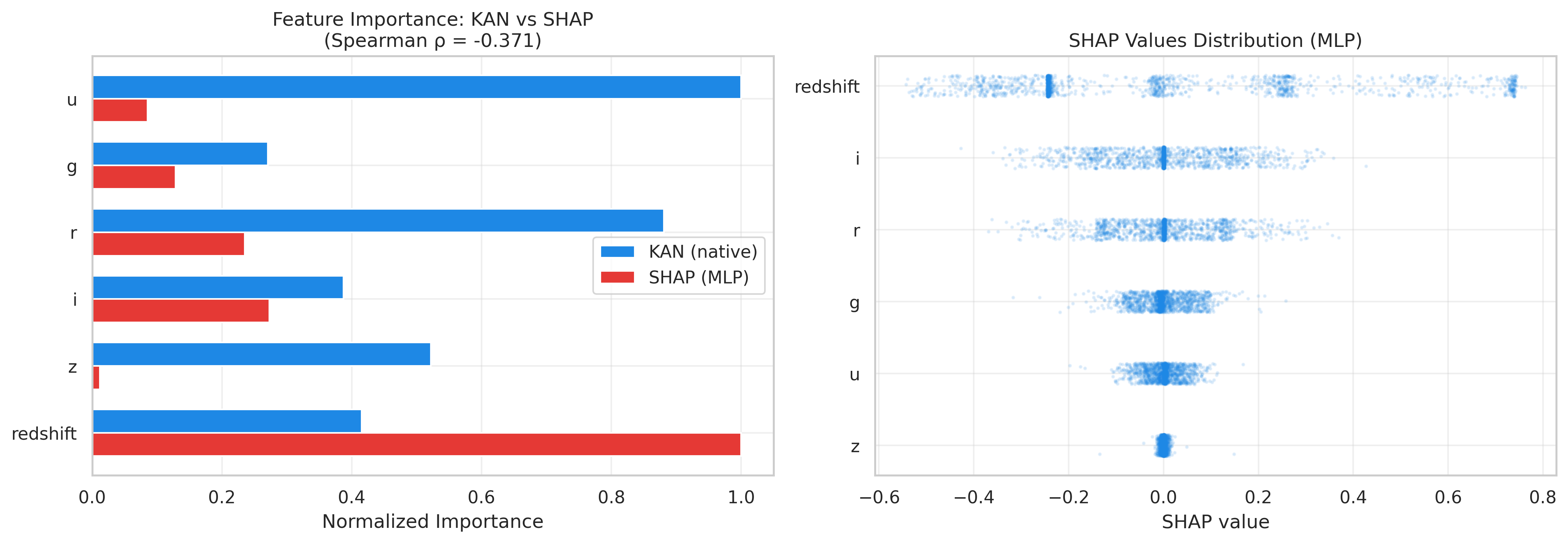}
\caption{Feature importance comparison: KAN native (activation magnitudes) vs.\ SHAP on MLP. Left: normalized importance. Right: SHAP value distribution. The two methods produce different rankings (Spearman $\rho = -0.37$), reflecting different attribution mechanisms.}
\label{fig:shap}
\end{figure}

KAN's native feature importance (first-layer activation magnitudes) is compared to SHAP~\cite{Lundberg2017} applied to MLP in Fig.~\ref{fig:shap}. KAN ranks the $u$-band first, then $r$ and $z$; SHAP on MLP ranks redshift first, then $i$ and $r$, with $u$ only fifth. The Spearman rank correlation is $\rho = -0.37$ ($p = 0.47$) --- no significant agreement.

The discrepancy reflects different representations rather than a failure of either method. KAN processes each feature through independent univariate splines, so its importance measures per-feature nonlinearity. MLP distributes information across shared hidden units, and SHAP captures marginal contributions including inter-feature interactions. Redshift dominates the SHAP ranking because it is a strong linear separator (stars vs.\ extragalactic objects); $u$-band dominates the KAN ranking because it carries the nonlinear structure needed for QSO identification.

\subsection{Probability calibration}
\label{sec:calibration}

\begin{figure}[ht]
\centering
\includegraphics[width=\textwidth]{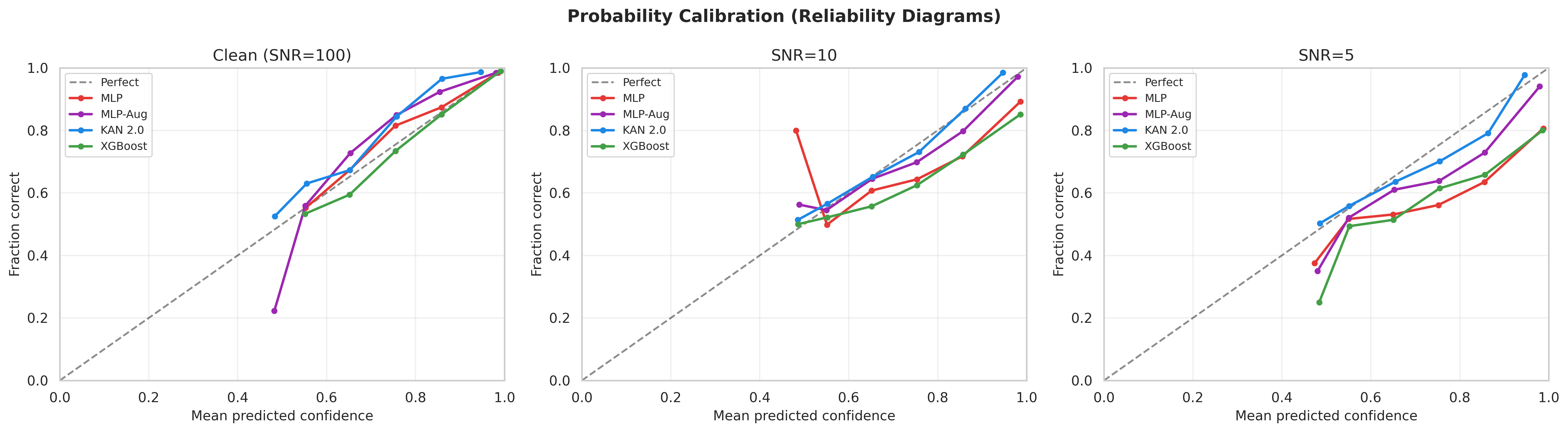}
\caption{Probability calibration (reliability diagrams) at three noise levels. All models become overconfident under noise; KAN and MLP-Aug degrade less than the baseline MLP.}
\label{fig:calibration}
\end{figure}

Reliability diagrams at three noise levels appear in Fig.~\ref{fig:calibration}. On clean data, XGBoost is best calibrated. Under noise (SNR\,$=$\,5), all models become overconfident, assigning high probabilities to increasingly incorrect predictions. KAN and MLP-Aug maintain better calibration than the baseline MLP at low SNR, in line with their smoother decision boundaries.

\subsection{Hybrid SNR-adaptive pipeline}
\label{sec:hybrid}

A hybrid pipeline routes predictions by model confidence: MLP-Aug handles samples where its predicted probability exceeds 0.85; uncertain samples go to KAN. At low SNR, 19--27\% of samples are routed to KAN (Table~\ref{tab:hybrid}). The hybrid improves accuracy at high SNR and at the lowest noise levels ($+$0.8~p.p.\ over MLP-Aug at SNR\,$=$\,1), though at intermediate noise (SNR\,$=$\,5) it slightly underperforms MLP-Aug alone (86.4\% vs.\ 86.9\%) --- KAN's lower clean accuracy dilutes the ensemble on the routed fraction.

\begin{table}[ht]
\centering
\caption{Hybrid pipeline accuracy (\%, Gaussian noise, 20 trials). Threshold\,$=$\,0.85. \%KAN: fraction of samples routed to KAN.}
\label{tab:hybrid}
\begin{tabular}{rcccc}
\toprule
SNR & MLP-Aug & KAN & Hybrid & \%KAN \\
\midrule
100 & 95.5 & 95.6 & \textbf{95.9} & 13.6 \\
10  & 91.7 & 89.9 & \textbf{90.6} & 15.5 \\
5   & \textbf{86.9} & 85.7 & 86.4 & 19.3 \\
3   & 81.1 & 81.7 & \textbf{82.2} & 23.2 \\
1   & 64.2 & 64.8 & \textbf{65.0} & 27.2 \\
\bottomrule
\end{tabular}
\end{table}

\begin{figure}[ht]
\centering
\includegraphics[width=\textwidth]{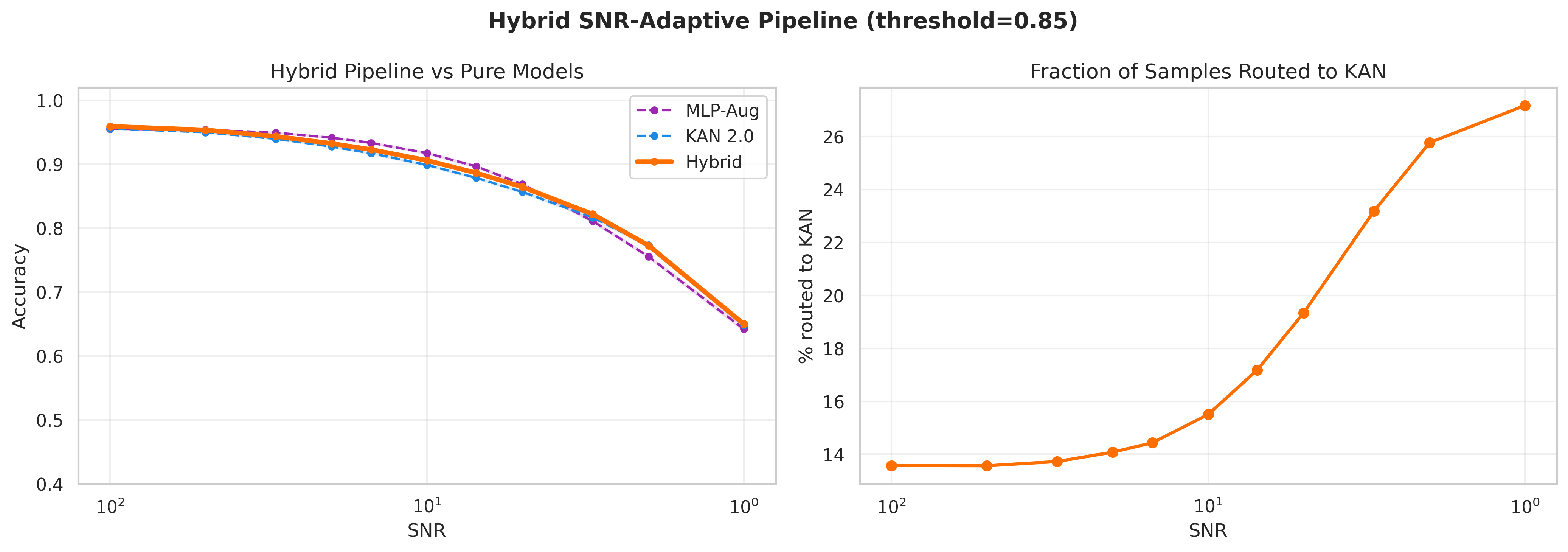}
\caption{Hybrid SNR-adaptive pipeline. Left: accuracy comparison. Right: fraction of samples routed to KAN increases as SNR decreases.}
\label{fig:hybrid}
\end{figure}

\subsection{Cross-dataset validation: DESI DR1}
\label{sec:desi}

The key experiments are repeated on an independent sample from DESI Data Release~1~\cite{DESI2024}: 100\,000 spectroscopically confirmed objects (balanced across the three classes, \texttt{ZWARN}\,$=$\,0). DESI provides different photometric bands --- Legacy Surveys $g$, $r$, $z$ plus Gaia BP and RP --- together with spectroscopic redshift (6 features).

\begin{figure}[ht]
\centering
\includegraphics[width=0.7\textwidth]{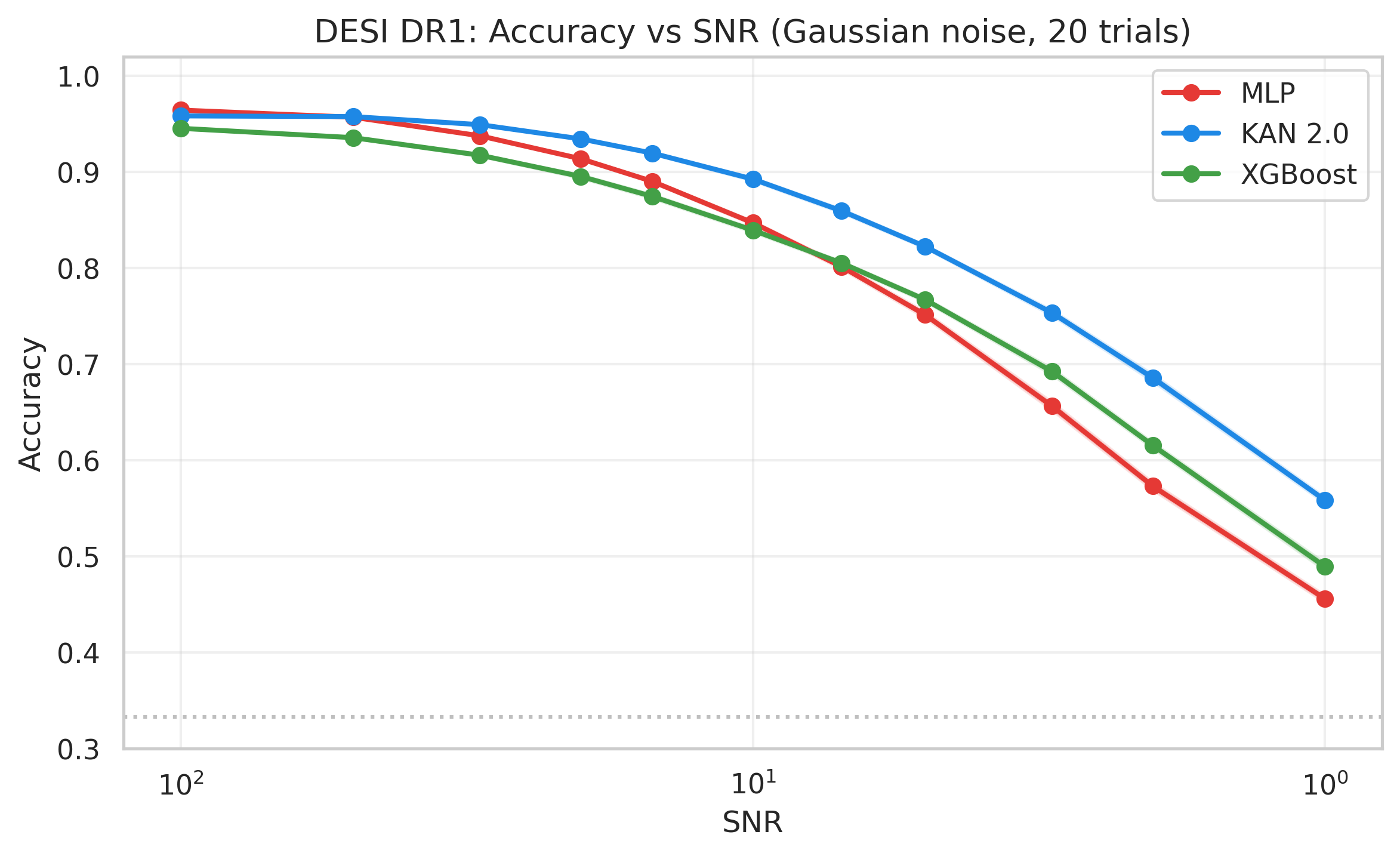}
\caption{DESI DR1: accuracy vs.\ SNR (Gaussian noise, 20 trials). The same pattern as SDSS: KAN appears superior in naive comparison.}
\label{fig:desi_snr}
\end{figure}

\begin{figure}[ht]
\centering
\includegraphics[width=0.7\textwidth]{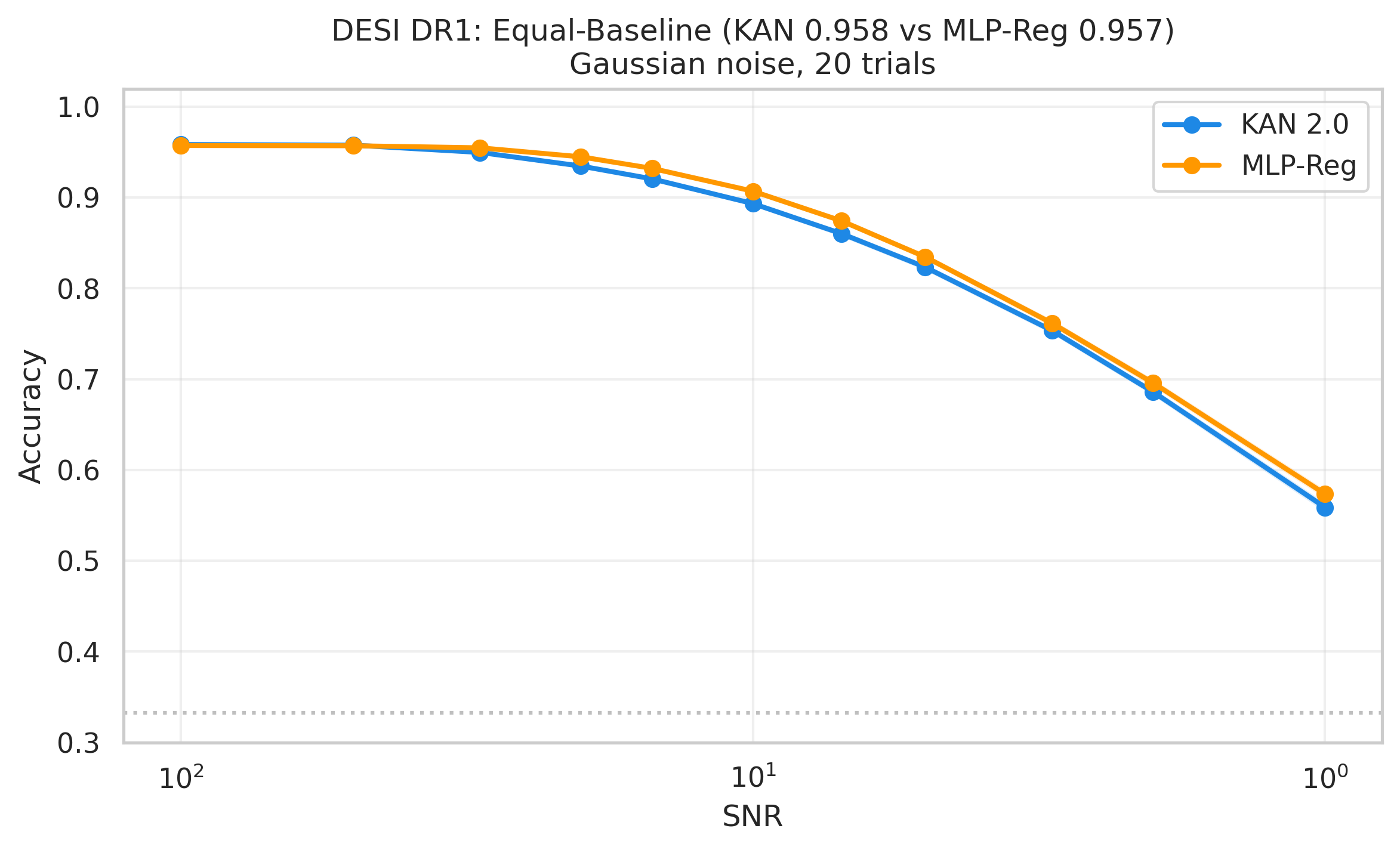}
\caption{DESI DR1 equal-baseline: MLP-Reg matches KAN's baseline and slightly outperforms it under noise, confirming the regularization explanation on an independent dataset.}
\label{fig:desi_equal}
\end{figure}

Baseline accuracies are comparable to SDSS (Table~\ref{tab:desi_baseline}). The naive comparison again favours KAN under noise (Fig.~\ref{fig:desi_snr}): at SNR\,$=$\,5, KAN retains 82.3\% versus 75.1\% for MLP ($+$7.1~p.p.). The equal-baseline experiment (Table~\ref{tab:desi_equal}, Fig.~\ref{fig:desi_equal}) again eliminates the gap: MLP-Reg (weight decay\,$=$\,0.003), matched to KAN's 95.8\% baseline, slightly \emph{outperforms} KAN at all noise levels (e.g., 83.5\% vs.\ 82.3\% at SNR\,$=$\,5). The regularization explanation holds across photometric systems.

\begin{table}[ht]
\centering
\caption{DESI DR1 baseline performance and noise robustness (Gaussian, 20 trials).}
\label{tab:desi_baseline}
\begin{tabular}{rccc}
\toprule
SNR & KAN & MLP & XGBoost \\
\midrule
100 & 95.8 & \textbf{96.5} & 94.6 \\
20  & 93.5 & 91.4 & 89.5 \\
10  & 89.3 & 84.7 & 83.9 \\
5   & 82.3 & 75.1 & 76.7 \\
3   & 75.3 & 65.6 & 69.2 \\
1   & 55.8 & 45.6 & 48.9 \\
\bottomrule
\end{tabular}
\end{table}

\begin{table}[ht]
\centering
\caption{DESI DR1 equal-baseline comparison (Gaussian, 20 trials). MLP-Reg tuned to match KAN's 95.8\% baseline.}
\label{tab:desi_equal}
\begin{tabular}{rcc}
\toprule
SNR & KAN~2.0 & MLP-Reg \\
\midrule
100 & 95.8 & 95.7 \\
20  & 93.5 & \textbf{94.5} \\
10  & 89.3 & \textbf{90.7} \\
5   & 82.3 & \textbf{83.5} \\
3   & 75.4 & \textbf{76.2} \\
1   & 55.9 & \textbf{57.4} \\
\bottomrule
\end{tabular}
\end{table}

\section{Discussion}
\label{sec:discussion}

\subsection{KAN as an implicit regularizer}
\label{sec:regularizer}

KAN's noise robustness is not an \emph{architectural} advantage but a \emph{regularization} effect. Three mechanisms are responsible:

(i)~\textbf{$C^2$ smoothness.} Cubic B-splines are twice continuously differentiable, suppressing high-frequency decision boundaries sensitive to input perturbations --- analogous to spectral regularization in kernel methods.

(ii)~\textbf{$L_1$ sparsity penalty.} The penalty on spline coefficients ($\lambda = 0.001$) reduces the effective number of active basis functions.

(iii)~\textbf{Univariate decomposition.} Each edge processes a single input dimension, so multi-dimensional noise is projected onto independent univariate learners rather than compounding across features.

Explicit weight decay on a standard MLP replicates this regularization almost exactly: in the equal-baseline experiment (Section~\ref{sec:controlled}, $\lambda_{\mathrm{wd}} = 0.005$), in the photometry-only setting (Section~\ref{sec:equal_noredshift}, $\lambda_{\mathrm{wd}} = 0.008$), and on DESI with different photometric bands (Section~\ref{sec:desi}, $\lambda_{\mathrm{wd}} = 0.003$). In the latter two cases MLP-Reg even slightly outperforms KAN. Existing MLP pipelines with proper regularization therefore do not need KAN for noise robustness.

\subsection{When KAN is still preferable}
\label{sec:when_kan}

KAN nonetheless has practical advantages:

\textbf{No tuning required.} KAN reaches near-optimal noise robustness with its default B-spline parameterization, without grid search over weight decay or noise augmentation ranges. For survey pipelines processing heterogeneous data where noise varies across the sky, this is convenient.

\textbf{Built-in interpretability.} The response functions (Fig.~\ref{fig:response}) directly show which features and feature ranges drive classification --- useful where understanding \emph{why} a source is classified matters. The $u$-band dominance (Fig.~\ref{fig:feature_importance}) matches two decades of colour-based classification work~\cite{Richards2002} and emerges without manual feature engineering. KAN's importance rankings differ from SHAP on MLP ($\rho = -0.37$, Section~\ref{sec:shap}), because the two methods capture complementary aspects: per-feature nonlinearity versus marginal contributions including interactions.

\textbf{Stronger robustness on colour indices.} On physically meaningful ratios (colour indices rather than raw magnitudes), KAN's smooth basis functions provide even stronger relative robustness (Section~\ref{sec:colors}). The larger baseline gap (5.4~p.p.) means that part of this advantage may again trace to implicit regularization; an equal-baseline experiment on colour indices would be needed to confirm a feature-space effect.

\textbf{Hybrid pipeline.} The confidence-based routing scheme (Section~\ref{sec:hybrid}) shows that MLP-Aug and KAN complement each other: MLP-Aug handles confident predictions efficiently, KAN improves accuracy on the uncertain, low-SNR tail.

\subsection{The cost of robustness}
\label{sec:cost}

KAN training takes ${\sim}$460\,s versus ${\sim}$24\,s for MLP (19$\times$ slower, single CPU core, Intel Xeon 2.2\,GHz). For large-scale surveys this limits KAN to selective application on faint sources or as an interpretability tool on representative subsamples.

\subsection{Limitations}
\label{sec:limitations}

(i)~\textbf{Synthetic noise.} Noise is injected into standardized features, not raw photon counts. Propagation through real photometric pipelines may differ.

(ii)~\textbf{Underfitting in the photometry-only setting.} KAN's 77.3\% baseline without redshift falls below MLP (82.7\%), indicating underfitting. The equal-baseline experiment (Section~\ref{sec:equal_noredshift}) confirms that MLP-Reg matches KAN even here, but the underfitting itself is a practical limitation.

(iii)~\textbf{Two photometric datasets.} Validation covers SDSS and DESI, both photometric classification tasks. The SDSS sample is class-imbalanced, the DESI sample balanced, which affects cross-dataset comparability. Generalization to imaging, spectroscopic, or time-series data is untested.

(iv)~\textbf{No equal-baseline test for colour indices.} The colour-index experiment (Section~\ref{sec:colors}) has a 5.4~p.p.\ baseline gap between KAN and MLP; an equal-baseline comparison was not performed.

(v)~\textbf{Symbolic regression was uninformative.} KAN~2.0's symbolic regression produced formulas with hundreds of terms (accuracy 73.4\%), offering no advantage over spline visualization. The feature may be more useful for lower-dimensional problems.

\section{Conclusions}
\label{sec:conclusions}

A controlled comparison of KAN~2.0 against MLP variants and XGBoost for stellar classification under noise yields the following conclusions.

(i)~KAN is not architecturally superior for noise robustness. A properly regularized MLP matches KAN's degradation curve to within 1~p.p.\ at all SNR levels --- on SDSS and DESI, with and without redshift, across different photometric systems. The 7--9~p.p.\ gap in naive comparisons traces to differences in effective regularization.

(ii)~KAN is a convenient implicit regularizer: it achieves near-optimal robustness out of the box, without hyperparameter search for weight decay or noise augmentation.

(iii)~Interpretability is KAN's genuine advantage. Learned response functions expose physically meaningful feature dependencies ($u$-band dominance, redshift threshold) that are inaccessible from MLP without post-hoc methods. KAN's native importance differs from SHAP ($\rho = -0.37$), as the two capture complementary aspects of the classification.

(iv)~Noise impacts object classes unequally: stars degrade fastest (F1: $0.97 \to 0.75$ at SNR\,$=$\,5), QSOs least (F1: $0.92 \to 0.91$), due to different levels of redundancy in classification cues.

(v)~Colour indices widen KAN's relative advantage over MLP, though an equal-baseline test would be needed to separate a feature-space effect from implicit regularization.

(vi)~A hybrid pipeline routing uncertain MLP predictions to KAN improves low-SNR accuracy while keeping MLP's efficiency at high SNR.

(vii)~Noise augmentation is the most effective single-model strategy when the noise distribution is known, but it requires choosing an SNR range matched to deployment conditions.

KAN is best suited to pipelines where noise characteristics are unknown or variable, where interpretability of feature mappings is scientifically valuable, and where the 19$\times$ training overhead is acceptable. For well-characterized noise, a noise-augmented MLP remains the most efficient choice.

\section*{Acknowledgements}

No specific funding was received for this work; no competing interests exist. Software used: \textsc{pykan}~\cite{Liu2024kan}, \textsc{PyTorch}~\cite{Paszke2019}, \textsc{scikit-learn}~\cite{Pedregosa2011}, \textsc{XGBoost}~\cite{Chen2016}, \textsc{SHAP}~\cite{Lundberg2017}.

\section*{Data Availability}

The SDSS DR17 stellar classification dataset is publicly available. DESI DR1 data are available at \url{https://data.desi.lbl.gov/public/dr1/}. Code and trained models: \url{https://github.com/dkrse/kan-vs-mlp-stellar-classification}.


\end{document}